\documentstyle[11pt,newpasp,twoside,epsf]{article}
\def\edcomment#1{\iffalse\marginpar{\raggedright\sl#1\/}\else\relax\fi}
\reversemarginpar

\begin{document}
\title{Tracing cosmic evolution with the Coma cluster}
 \author{Bianca M. Poggianti}
\affil{Osservatorio Astronomico, vicolo dell'Osservatorio 5, 35122 Padova -I}
\author{Terry J. Bridges}
\affil{Anglo-Australian Observatory, PO Box 296, Epping, NSW 1710, Australia}
\author{Bahram Mobasher}
\affil{Space Telescope Science Institute, 3700 San Martin Drive, Baltimore, MD 21218, USA}
\author{Dave Carter}
\affil{Liverpool john Moores University, Twelve Quays House, Egerton Wharf, Birkenhead, Wirral, CH41 1LD, UK}

\begin{abstract}
We summarize here the first results of a new spectroscopic survey of galaxies in the Coma
cluster with $-20.5<M_B<-14$. Differences between the last epoch of star formation (SF) activity
in S0 and elliptical galaxies are discussed. Furthermore, we show that about half of
all galaxies without present-day SF display signs of activity at $z<1-2$. 
Systematic trends of most recent SF epoch with galaxy luminosity are presented.  
\end{abstract}
\vspace{-0.1in}
\section{Introduction}
In this contribution we report on the first results of a new spectroscopic survey of galaxies
in the Coma cluster. This survey was designed to have three main characteristics:\newline
1) It extends over almost 7 mag in galaxy luminosity ($-20.5<M_B<-14$), well into the dwarf
regime.  In fact, this is by far the largest spectroscopic sample of cluster dwarf galaxies
to date.\newline
2) It covers two areas of $33' \times 50'$ ($1 \times 1.5$ Mpc) 
towards the center and the southwest region
of Coma. The photometry (B and R) has been obtained with the Japanese Mosaic CCD camera 
at the William Herschel Telescope (WHT) in collaboration with M. Doi, M., Iye, N.Kashikawa,
Y. Komiyama, S. Okamura, M. Sekiguchi, K., Shimasaku, M. Yagi, N. Yasuda.\newline
3) It is essentially a magnitude limited sample, without significant color or morphological
bias, therefore it includes galaxies of all Hubble types.\newline
We obtained multifiber spectra with the WYFFOS fibre spectrograph at the WHT
for about 300 galaxies members of Coma and 200 field galaxies.
The spectra cover the central 2.2 arcsec (1.3 kpc) of the galaxies and have a mean
S/N ranging from 16 for the brightest subset to 8 for the faintest subset. Hence,
the errorbars on the spectral line measurements are large for the faint galaxies, and 
for them only broad conclusions regarding ages and metallicities can be reached.
\begin{figure}
\plottwo{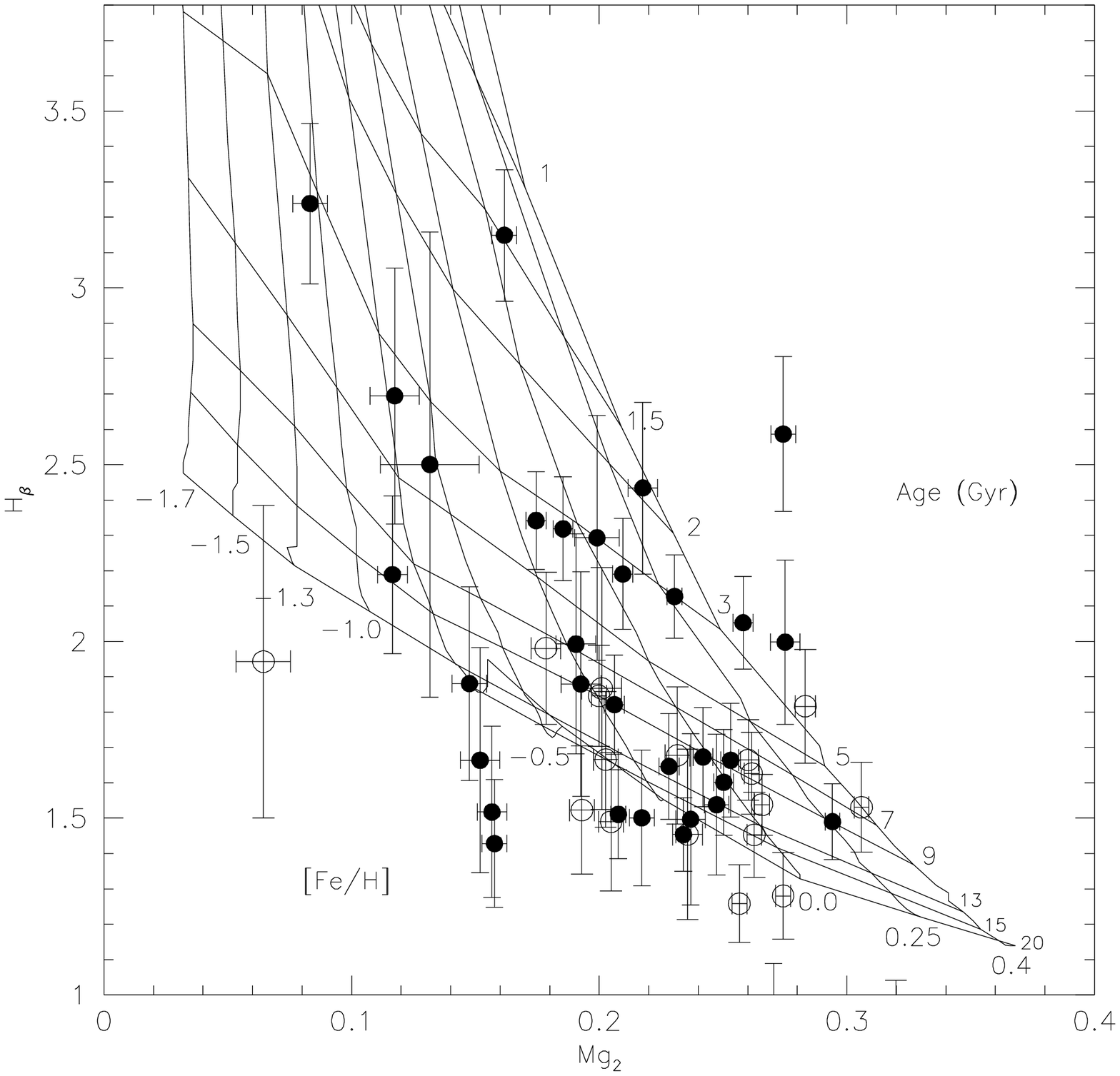}{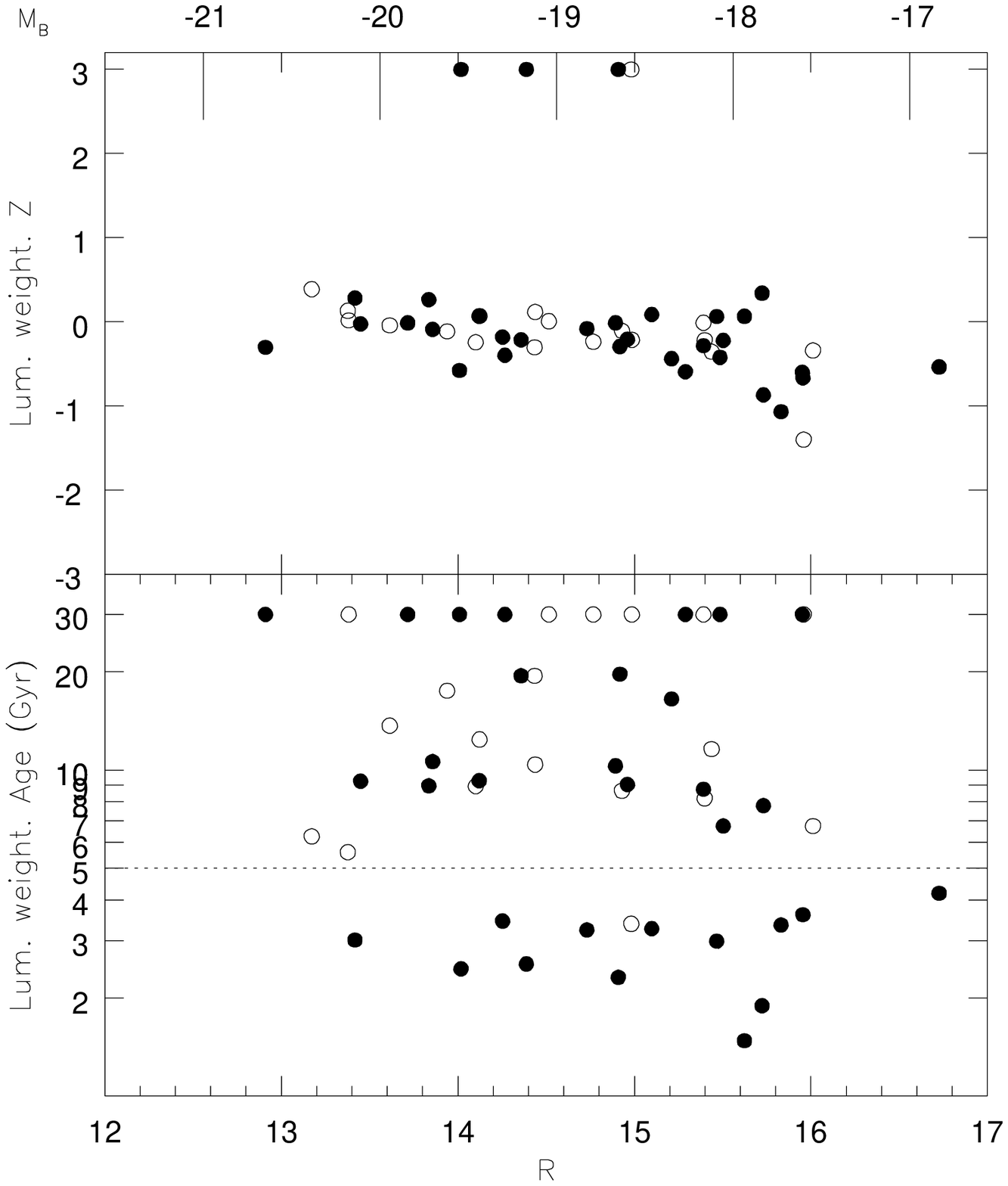}
\caption{{\it Left.} $\rm H\beta$ versus $\rm Mg_2$ index strength for ellipticals
(empty circles) and S0 galaxies (filled circles).
{\it Right.} Luminosity weighted metallicities (top) and ages (bottom)
of ellipticals (empty circles) and S0 galaxies (filled circles) as a function
of the R-band magnitude.}
\end{figure}
{\it Here we consider only the $\sim 250$ galaxies without emission lines in their spectra.}
For them, we have measured the absorption line indices in the Lick system. The stellar 
properties (luminosity weighted ages and metallicities) 
have been derived from index-index diagrams, comparing with spectrophometric
models based on the Padova isochrones (http://astro.sau.edu/\~{}worthey/dial/dial\_a\_pad.html,
Worthey). 
A full description of the results presented here can be found in Papers III and IV.
The dataset is described in Papers I and II.
\vspace{-0.2in}
\section{S0 versus elliptical galaxies}
There are 19 ellipticals and 33 S0s in our sample with a secure morphological
classification from Dressler (1980). All but one of the ellipticals are consistent
with luminosity weighted ages $\geq 9$ Gyr, while 13 S0s have ages $< 5$ Gyr (Fig.~1, left).
Thus, a significant fraction ($> 40$\%) of the S0s show evidence for some recent SF.
This is consistent with the hypothesis that they evolved from star-forming spirals
that were accreted onto the cluster and had their SF truncated at intermediate redshifts
(Dressler et al. 1997, Fasano et al. 2000, see also van Dokkum and Treu in these proceedings).
This difference in age spread between Es and S0s 
is similar to the results of Kuntschner \& Davies (1998) in
the Fornax cluster and Smail et al. (2001) in A2218.
In contrast, other works (e.g. Jones, Smail \& Couch 2000, Ellis et al. 1997) have not found
a statistically significant difference between the ages of the stars in S0s and Es.
It is obviously important to understand whether S0s with recent SF are a {\it common} phenomenon
in clusters or not. In this respect, it is interesting to look at the age distributions
of galaxies as a function of magnitude.\footnote{We note that S0s and Es follow a 
broadly similar metallicity-luminosity relation (top in the right panel of Fig.~1), although the
relation for the S0s is slightly steeper and more scattered than that of the Es.}
\begin{figure}
\vspace{-0.2in}
\plotone{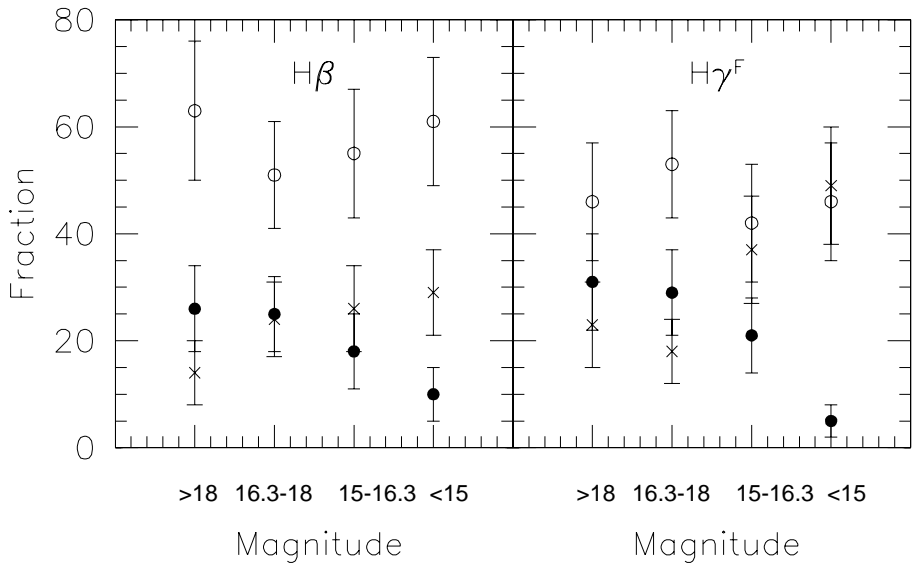}
\vspace{-0.2in}
\caption{Fraction of young (filled dots, age $<3$ Gyr), 
intermediate-age (crosses, 3 to 9 Gyr) and old (empty dots, age $>9$ Gyr) 
galaxies within each magnitude bin as
derived from the $\rm H\beta$/$\rm Mg_2$ diagram (left) and the 
$\rm{H\gamma}_F$/$<$Fe$>$ diagram (right).}
\end{figure}
In Fig.~1 (right), the population of S0s with recent SF is noticeable 
below the dotted line: only 4 of the 
``young'' S0s are brighter than $M_B = -19$. In fact, considering galaxies brighter and fainter
than $M_B = -19$, bright and faint Es have similar age distributions, while the fraction
of S0s with recent SF is higher at fainter luminosities. Therefore, a possible reason
for the discrepant results found so far (recent SF in cluster S0s, yes or no) could be
the luminosity range explored by the different studies. Given the luminosity
function of spirals at $z \sim 0.5$ and the expected fading if their SF is halted,
the great majority of the spirals evolved into S0s should be typically fainter
than $M_B = -20/-19$.\footnote{Those S0s with recent SF have an asymmetric distribution 
in the sky, being
preferentially located in a region east/north-east of NGC4874 (the cluster centre), 
while older S0s and ellipticals 
are spread out both north and south of NGC4874.} 
\vspace{-0.2in}
\section{The infall history of the Coma cluster}
In the following we will consider the complete spectroscopic sample with no emission
lines, including galaxies of all magnitudes and Hubble types. This sample 
represents the total cluster population of galaxies without current SF.\newline
There is a broad range of ages, from younger than 3 Gyr to older than 9 Gyr, 
among galaxies of any magnitude. However, there are systematic trends of age 
with galaxy magnitude. Dividing all galaxies into four magnitude bins,
$\sim 50$ \% of galaxies in any bin are ``old'', i.e. with no sign of SF during the last 9 Gyr
(at $z<1-2$). 
As shown in Fig.~2, the proportion of {\it luminous} galaxies that are passive at z=0
and that experienced a SF activity at intermediate redshift ($0.35<z<1$) is 
30 to 50 \%,
and it is higher than the fraction of {\it dwarf} galaxies that were
active at that epoch. This is consistent with the observations of 
current/recent SF in a large number of
galaxies in intermediate-z clusters (e.g. Dressler et al. 1999). 
At low redshift, instead, SF involved a higher fraction of faint than bright galaxies.
These results can be used to trace the infall history of galaxies onto Coma, and the
consequent star formation history of its present-day galaxies.
The luminosity(mass) dependence of the most recent SF
epoch points to a ``down-sizing effect'', and can 
be compared with the accretion
history of clusters predicted by cosmological models. 

\end{document}